\documentclass[useAMS,usenatbib,usegraphicx,referee]{mn2e}
\usepackage{graphicx,amssymb,amsmath,times,verbatim}
\usepackage{color}

\usepackage{amssymb,amsmath}
\usepackage[normalem]{ulem}

\def\lta{\mathrel{\spose{\lower 3pt\hbox{$\mathchar"218$}}
     \raise 2.0pt\hbox{$\mathchar"13C$}}}
\def\gta{\mathrel{\spose{\lower 3pt\hbox{$\mathchar"218$}}
     \raise 2.0pt\hbox{$\mathchar"13E$}}}

\def\mathnew{\mathsurround=0pt}

\def\simov#1#2{\lower .5pt\vbox{\baselineskip0pt \lineskip-.5pt
\ialign{$\mathnew#1\hfil##\hfil$\crcr#2\crcr\sim\crcr}}}

\newcommand\apj{ApJ}%
\title[]{GAMMA RAY FLARE OF PKS 1222+216 IN 2010: EFFECT OF JET DYNAMICS AT THE RECOLLIMATION ZONE}
\author[P. Kushwaha et al.]{Pankaj Kushwaha$^{1}$\thanks{E-mail:pankaj563@tifr.res.in}, 
			  S. Sahayanathan$^{2}$\thanks{E-mail:sunder@barc.gov.in}, 
			   Lekshmi Resmi$^{3}$, K. P. Singh$^{1}$,
\newauthor Sudip Bhattacharyya$^{1}$ \& Dipankar Bhattacharya$^{4}$\\
$^{1}$ Department of Astronomy \& Astrophysics, Tata Institute of Fundamental Research, Mumbai, India \\
$^{2}$ Astrophysical Sciences Division, Bhabha Atomic Research Centre, Mumbai, India\\
$^{3}$ Indian Institute of Space Science \& Technology, Thiruvananthapuram, India \\
$^{4}$ Inter-University Center for Astronomy \& Astrophysics, Pune, India}

\begin{document}

\maketitle

\begin{abstract}
The $\gamma$-ray flare of PKS 1222+216, observed in June 2010, is interpreted as an outcome of jet 
dynamics at recollimation zone. We obtained the $\gamma$-ray light-curves in three different energy 
bands, namely, 100--300 MeV, 300 MeV--1 GeV and 1--3 GeV from observations by the \emph{Fermi} 
Large Area Telescope (LAT). We also use the \emph{Swift}--XRT flux from 0.3--10 keV obtained from
archival data. We supplement these with the 0.07--0.4 TeV observations with MAGIC telescope, available 
in the literature. The detection  of source at very high energy (VHE, $E>100$ GeV) with a differential photon 
spectral index of $2.7\pm0.3$ and the rapid variability associated with it suggests that the emission arises 
from a compact region located beyond the broad line emitting region. The plausible $\gamma$-ray emission 
mechanism can then be inverse Compton scattering of IR photons from obscuring torus. Further, the decay 
time of LAT flare cannot be explained by considering simple radiative loss mechanisms. 
Hence, to interpret the LAT light curves, we develop a model where the broadband emission originates from 
a compact region, arising plausibly from the compression of jet matter at the recollimation zone. The flare is 
then expressed as an outcome of jet deceleration probably associated with this focusing effect. Based on 
this model, the rise of the LAT flare is attributed to the opening of emission cone followed by the decay resulting 
from jet deceleration. The parameters of the model are further constrained by reproducing the broadband spectral 
energy distribution of the source obtained during the flare episode. Our study suggests that the particle energy density 
exceeds magnetic energy density by a large factor which in turn may cause rapid expansion of the emission region.
However, near equipartition can be achieved towards the end of LAT flare during which the compact emission region
would have expanded to the size of jet cross-section.
\end{abstract}


\section{Introduction} \label{sec: I}
Flat spectrum radio quasars (FSRQs) are radio loud active galactic nuclei (AGNs) with relativistic jet 
oriented close to the line of sight of the observer. Under the unification theory, they are classified along with 
BL Lacs as blazars \citep{1995PASP..107..803U}. Non-thermal emission extending from radio to $\gamma$-rays, rapid 
flux variability and high degree of polarization are some of the common properties observed in blazars 
\citep{2000AIPC..515...19S,2008PASJ...60..707F,2004NewAR..48..367K}. Their spectral energy distribution 
(SED) is characterized by a typical double-humped feature extending from radio to $\gamma$-ray energies 
\citep{1999ApJ...514..138K,1998ApJ...497..178W} and in a few cases, up to  very high energies (VHEs, E$>$100 GeV) 
\citep{2008Sci...320.1752M, 2010HEAD...11.2706W, 2011ApJ...730L...8A}. The observed rapid flux variability
suggests the emission to arise from a compact region located close to the central engine and moving down the jet at 
relativistic speed \citep{1995MNRAS.273..583D}. The low energy emission, extending from radio to UV/X-ray, is 
generally interpreted as synchrotron emission from a non-thermal population of electrons, while the high energy 
emission is believed to originate from inverse Compton scattering of low energy photons. 
The target photons for inverse Compton scattering can be the synchrotron photons from the jet themselves (SSC) 
\citep{1981ApJ...243..700K, 1985ApJ...298..114M, 1989ApJ...340..181G} or photons external to the jet (EC) 
\citep{1987ApJ...322..650B, 1989ApJ...340..162M, 1992A&A...256L..27D}. Simple emission models that consider only 
synchrotron and SSC processes, cannot explain the $\gamma$-ray emission from FSRQs and one needs to invoke EC emission 
to explain the SED satisfactorily \citep{2001ApJ...553..683H, 2009ApJ...703.1168B, 2012MNRAS.419.1660S}. A few plausible 
target photons for the EC process are photons from the accretion disk \citep{1993ApJ...416..458D, 1997A&A...326L..33B}, 
reprocessed accretion disk photons from the broad line emitting region (BLR) \citep{1994ApJ...421..153S, 1996MNRAS.280...67G} 
and/or infra-red (IR) photons from the dusty obscuring torus \citep{1994ApJ...421..153S, 2000ApJ...545..107B, 2008MNRAS.387.1669G}.

PKS 1222+216 ($z=0.432$) is a flat spectrum radio quasar detected at VHE by the
Major Atmospheric Gamma-ray Imaging Cherenkov Telescope \citep[\emph{MAGIC},][]{2010ATel.2684....1M,2011ApJ...730L...8A}.
It is the third FSRQ detected at VHE after 3C 279 and PKS 1510-089 
\citep{2008Sci...320.1752M, 2010HEAD...11.2706W}. PKS $1222+216$ has been active at LAT energies
since September 2009, undergoing occasional brightness enhancements \citep{2011ApJ...733...19T}. 
Such flaring episodes were also detected in other observatories operating at different/similar wavebands 
\citep{2010ATel.2641....1B,2010ATel.2626....1C}. The source underwent two major flares of $\sim 10^{-5}$ph 
cm$^{2}$ s$^{-1}$ ($0.1-300$ GeV, $>10\sigma$) in April and June 2010 \citep{2011ApJ...733...19T}. The second 
flare in June was associated with a rapid VHE flare observed by the \emph{MAGIC} telescope 
(on June 17, 2010), with a flux doubling timescale of $\sim$ 10  min 
\citep{2010ATel.2684....1M,2011ApJ...730L...8A}. The \emph{Swift}--XRT did not cover the peak of the flares in April or 
June, but followed the source in the decaying part of the June flare \citep{2011A&A...534A..86T}. We have analysed the 
LAT data of PKS 1222+216 from June 16th to 22nd in three different energy bands and obtained the light-curves of the
flare. We have also analysed the contemporaneous \emph{Swift}--XRT data to obtain a time-averaged broad-band spectrum (\S 2).

The observed VHE spectrum with a differential photon spectral index of $2.7\pm0.3$, and the rapid variability 
 introduces additional constraint on the location and the size of the emission region \citep{2012MNRAS.425.2519N}. 
The observed VHE spectral index suggests that the inverse Compton process happens in the Thomson 
regime as scattering in Klein-Nishina regime predicts a steeper spectrum \citep[$\sim 4$;][]{2009herb.book.....D}. 
This constraint rules out the possibility of EC scattering of BLR photons and hence demands the emission region 
to be located beyond the BLR \citep{2009MNRAS.397..985G, 2012MNRAS.425.2519N}. On the other hand, the rapid 
variability timescale of $\sim$10 min requires a smaller emission region compared to the jet cross-section at this 
distance. This led \citet{2011A&A...534A..86T} to propose a blob-in-jet model where the high energy emission 
originates from a compact region buried inside the jet along with the standard emission region covering the whole 
jet cross-section. They further argued that such a scenario is possible when an expanding jet interacts with the 
external medium resulting in a recollimation shock, which in turn compresses the matter towards the jet axis, 
giving rise to a compact emission region. The dynamics of the outflow at the recollimation zone have been studied 
by \citet{2009ApJ...699.1274B} following a semi analytical approach and they found that the focussing of jet due to 
the recollimation shock is also associated with the deceleration of jet flow. Similar result of jet deceleration at 
recollimation shock has also been seen in high resolution numerical simulations \citep{2007MNRAS.382..526P}. 
Deceleration of jet can be understood as a result of radiative losses \citep{1999APh....11...19M} and/or due to 
sweeping up of ambient/jet matter \citep{2007MNRAS.382..526P,2009ApJ...692.1374B,1999ApJ...512..699C}. If the jet axis is aligned 
close to the line of sight of the observer, then deceleration of jet will result in a time dependent Doppler boosting 
with an increase in the opening angle of the emission cone. 

In this paper, we interpret the $\gamma$-ray light curve of PKS 1222+216 during the flare on June 2010 as a 
result of jet dynamics happening at the recollimation zone. The kinetic equation describing the evolution of
 electron spectrum in the emission region is solved numerically. The resultant photon spectrum is obtained by 
convolving the time dependent electron distribution with single particle emissivity corresponding to various
emission processes. The rise of photon flux during the flare results from an increase in the opening angle of the 
emission cone, while the fall is governed by the effects of jet deceleration.

In the next section, we describe the data analysis procedures, and in \S \ref{sec:obs_con} we describe the constraints 
derived from the observations and the rationale behind the present model. In \S4, we present the details of the model 
and the underlying assumptions. Finally in \S \ref{sec:resndis}, we discuss the results obtained followed by conclusions in 
\S\ref{sec:conclude}. A cosmology with $\Omega_m = 0.3$, $\Omega_\Lambda = 0.7$ and $H_0 = 70\;km\;s^{-1}\;Mpc^{-1}$ is used 
in this work which corresponds to a luminosity distance $d_L$= $2.37$ Gpc for $z=0.432$.

\section{Data analysis}
The Large Area Telescope (LAT) on board the {\emph{Fermi Gamma-ray Space Telescope}} is sensitive to $\gamma$-rays 
from $20$~MeV to $>300$~GeV \citep{2009ApJ...697.1071A}. The entire sky is covered in its normal scanning operation mode 
in $\sim 3$ hours, thereby the time evolution of the $\gamma$-ray sources is regularly monitored. Automated follow-ups 
for \emph{Fermi}-LAT sources at lower wavelengths are conducted with various instruments and observatories. For example, 
the \emph{Swift}--XRT monitors any source that has flared above $10^{-6}$~ph cm$^{-2}$ s$^{-1}$ in LAT. In this paper, 
we have used the publicly available $\gamma$-ray and X-ray data from LAT and XRT respectively, during the $\gamma$-ray 
flaring period of PKS 1222+216 (MJD: 55363-55370).

{\emph{LAT data analysis:}} LAT data between 16th-22nd June 2010 (MJD: 55363-55370) were analysed using 
\emph{Fermi}-LAT Science tool version v9r27p1 which was the latest publicly available release during the time of analysis. 
Only the events that have energy  $>100$~MeV and zenith angles $<$ 100$^\circ$ were considered to avoid calibration 
uncertainties and earth limb. We selected the events tagged as `source class events' (evclass 2) in the {\textit{photon data}}, 
and good time intervals using the logical expression ``(DATA\_QUAL==1)\&\&(LAT\_CONFIG==1)\&\&ABS(ROCK\_ANGLE)$<52$''.  
We followed `unbinned likelihood analysis' \citep{1996ApJ...461..396M} (python implementation of \emph {gtlike}) to model 
photons from a region of interest (ROI) of $15^\circ$ centred around the position of the blazar. The exposure map was 
created for the ROI plus a $10^\circ$ annulus to take care of photons from outside the ROI possibly entering the region 
due to the large PSF. We used the \emph{pass 7} (\emph{P7SOURCE\_V6}) instrument response function along with 
galactic diffuse emission model (gal\_2yearp7v6\_v0.fits) and isotropic background model (iso\_p7v6source.txt). Point 
sources in the region were selected and modelled based on the LAT second-year catalogue 
\citep[gll\_psc\_v08.fit;][]{2012ApJS..199...31N}\footnote{PKS 1222+216 is fitted with a log parabola model}. The typical 
systematic uncertainties in the flux values are 10\% at 100 MeV, and 5\% between 316 MeV to 10 GeV.

\begin{figure}
\begin{center}
 \includegraphics[scale=1,angle=0]{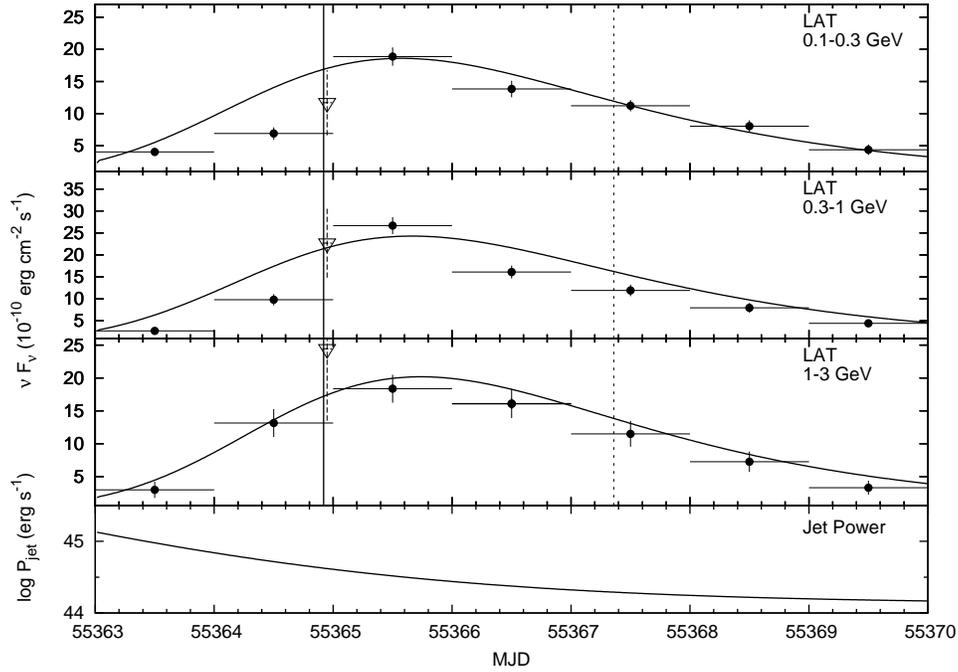}
\end{center}
\caption{The daily binned \emph{Fermi}-LAT light-curves of PKS 1222+216 in three energy bands obtained during the flare along 
with the model light curves. The instantaneous jet power ($P_{jet}$) is shown in the bottom panel. Error bars are standard 
1$\sigma$, statistical only. The estimated systematic uncertainties in the fluxes are 10\% at 100 MeV, and 5\% between 316 
MeV to 10 GeV. The solid and dashed vertical line marks the epoch of VHE and XRT observation during the LAT flaring episode. 
Inverted triangles are LAT fluxes at the time of VHE detection by \emph{MAGIC}, extracted from a two hour integrated LAT 
data (see \S 2).}
\label{fig:lc}
\end{figure}

We thus obtained the daily $\gamma$-ray light-curves in three energy bands ($100$ - $300$~MeV, $300$ - $1000$~MeV, 
and $1000$ - $3000$~MeV) following the above procedure and shown in Figure \ref{fig:lc} (filled circles). Point sources 
with test statistics (TS) values $\leq 0$ were removed from the source model during analysis. Fluxes from a two hour 
binned LAT data (Fig. \ref{fig:lc}, inverted triangles) were also obtained around MJD 55364.9, an epoch encompassing 
the VHE detection by the \emph{MAGIC} (Fig. \ref{fig:lc}, solid vertical line) using the best fit model parameters obtained 
for $0.1-300$ GeV. A significance criteria of $3\sigma$ corresponding to a TS value of $\sim10$ was used for the 
source detection.

{\emph{Swift}--XRT analysis:} We have used XRT data from June 20th (MJD $55367.4$), soon after the end of the flare, 
for creating quasi-simultaneous SEDs as shown in Figure \ref{fig:spec}. The XRT observation of the second flare 
started on June 20th, and thus missed the peak of the LAT flare and the epoch of the MAGIC observation. To create the 
SED we used the {\textit{HEASOFT}} package (version 6.13) and followed the standard procedure for XRT data analysis by 
obtaining source spectrum from a region of $90\%$~PSF ($\sim 47''$, \citet{2005SPIE.5898..360M}) and a background 
spectrum from nearby uncontaminated regions. We modeled the spectra as a power-law multiplied by an absorber (phabs) 
in {\emph{XSPEC}}. We fixed the neutral hydrogen column density at the Galactic value of $N_H=2.12\times10^{20}$cm$^{-2}$ 
obtained by the LAB Galactic HI survey \citep{2005A&A...440..775K}. The SED points were extracted using the best fit model 
parameters ($\Gamma_p = 1.60\pm0.11, dN/dE\sim E^{-\Gamma_p}$) and subsequently corrected for Galactic absorption.

Finally, the IR-Optical-UV data were obtained from literature \citep{2011A&A...534A..86T,2011ApJ...732..116M,2010ATel.2626....1C}, 
and the SED of the source was constructed for two epochs: one for simultaneous LAT-VHE observation and second 
for simultaneous LAT-XRT observation. Both the SEDs are shown in Fig. \ref{fig:spec} (black and grey data points 
respectively) and the corresponding epochs are demarcated by vertical lines (solid and dashed respectively) in 
Fig. \ref{fig:lc}. The LAT spectrum of the first SED is extracted using two hours of observed data whereas 
for the latter, we have used six hours of observed LAT data.

\section{Observational Constraints}\label{sec:obs_con}
The $\gamma$-ray emission from FSRQs is generally attributed to the EC scattering process. The plausible
target photons for the EC scattering are the IR photons from the dusty torus and/or the dominant Lyman-$\alpha$ 
photons from the BLR. However, scattering of BLR photons to VHE will happen in Klein-Nishina regime 
\citep{2009MNRAS.397..985G,2011ApJ...730L...8A, 2012MNRAS.425.2519N, 2012MNRAS.419.1660S} resulting in a
 steeper photon spectra \citep[$\sim 4$;][]{2009herb.book.....D}, in contrast with the observed VHE photon spectral 
index of $2.7\pm0.3$ \citep{2011ApJ...730L...8A}. This suggests that the VHE emission is due to EC scattering of IR 
photons which happens in the Thomson regime. This in turn demands the emission region to be located beyond the BLR 
and this condition along with the observed rapid variability (doubling time $\sim$ 10 min) decides the location 
and the size of the emission region \citep{2012MNRAS.425.2519N}. Incidentally, the condition on the target photons 
for EC scattering to happen in the Thomson regime also assures the transparency of VHE photons against the pair 
production losses \citep{2009MNRAS.397..985G,2012MNRAS.419.1660S}.

The IR bump present in the SED of PKS 1222+216 (Fig. \ref{fig:spec}) can be readily reproduced by a blackbody 
spectrum at temperature 1200 K \citep{2011ApJ...732..116M} and luminosity $L_{IR} \simeq 1\times10^{46}$ 
erg s$^{-1}$. For the sake of simplicity, if we assume this emission to originate from a spherical shell, then its 
location from the central engine corresponds to $R_{IR}\simeq 7\times10^{18}$ cm, with a disk luminosity of  
$L_{UV}\simeq 7\times10^{46}$ erg s$^{-1}$ (obtained from the observed UV flux) and a covering 
factor of 0.15 ($\approx L_{IR}/L_{UV}$) \citep{2013MNRAS.433.2380K}. For a conical jet geometry,  
with opening angle $\phi\approx 0.008(\Gamma/32)^{-1} $ radian \citep{2009A&A...507L..33P} where $\Gamma$
is the jet bulk Lorentz factor, the jet cross-section at this distance will be $D \approx 6 \times10^{16}$ cm. 
On the other hand, the constraint on the proper size of emission region, derived from the VHE variability is given 
by\footnote{Quantities with prime are measured in the rest frame of the emission region.}
\begin{align} \label{eq:VHEvar}
 R^\prime&\lesssim \frac{\delta}{(1+z)}c\,t_{var} \nonumber \\
   &\simeq 3\times10^{14} \left(\frac{\delta}{22}\right) \left(\frac{t_{var}}{10~min}\right)~\text{cm}
\end{align}
where $\delta$ is the Doppler factor given by $\delta=[\Gamma (1-\beta\, cos\,\theta)]^{-1}$ with
$\beta$ being the dimensionless jet velocity and $\theta$ is the angle between the jet axis and line of sight of the observer.
However, this size is much smaller compared to the jet cross-section, thereby demanding stringent constraints on
the modelling of broadband spectrum. A possible explanation for such a
compact emission region can be the compression of jet matter by a recollimation
shock \citep{2009ApJ...699.1274B,2011ApJ...730L...8A,2011A&A...534A..86T}.

Further, if we consider the observed GeV emission 
($E_{ph}=\delta \Gamma \gamma^2 \epsilon_0/(1+z)$; \citet{2013MNRAS.433.2380K}) as a result of inverse Compton 
scattering of IR photons ($\epsilon_0=2.82kT$), then 
the Lorentz factor ($\gamma$) of the electron responsible for this emission can be estimated as
\begin{align}
	\gamma^\prime_{1GeV} &= \sqrt{\frac{E_{ph}}{\epsilon_0} \frac{(1+z)}{\delta \Gamma}}	\nonumber \\
			    &\approx 2.6\times10^3\left(\frac{E_{ph}}{1~GeV}\right)^{1/2}  \left(\frac{T}{1200~K}\right)^{-1/2} 
		      \left(\frac{\delta}{22}\right)^{-1/2} \left(\frac{\Gamma}{32}\right)^{-1/2}
\end{align}
where $E_\gamma$ is the energy of the observed $\gamma$-ray photon and $\epsilon_0$ is the seed
photon energy in the AGN frame.
The corresponding cooling time-scale ($t_{cool}=\gamma/\dot{\gamma}$) of the electron responsible for GeV emission
in observer's frame will then be
\begin{align} \label{eq:cool}
	t_{cool,1GeV} &\approx \left(\frac{1+z}{\delta}\right)\left(\frac{3m_e c}{4\sigma_T\gamma^\prime_{1GeV} u_{ir}^\prime}\right)	\nonumber \\
		      &\approx 0.8 \left(\frac{T}{1200~K}\right)^{-7/2}
	\left(\frac{\delta}{22}\right)^{-1/2} \left(\frac{\Gamma}{32}\right)^{-3/2}  ~\text{min}
\end{align}
where $c$, $\sigma_T$, $m_e$ are the speed of light, Thomson scattering cross-section, electron rest mass respectively
and  $u_{ir}^\prime$ is the IR photon energy density in the emission region frame, related to the IR photon energy density
in the AGN frame by
\begin{equation}
 u_{ir}^\prime \approx \Gamma^2 \frac{4\sigma}{c}   T^4
\end{equation}
Here $\sigma$ is Stefan-Boltzmann constant. However, this time is too short compared to the decay time of 
the $\gamma$-ray flare ($\sim 3$ days) observed by LAT. 
Hence, the flaring behaviour of PKS 1222+216 cannot be attributed to a simple electron cooling 
mechanisms but possibly an outcome of jet dynamics \citep{2009ApJ...692.1374B}. If we interpret
the flaring activity as a result of efficient particle acceleration at recollimation zone, then the 
decay of flare can be related to the deceleration of jet flow due to radiative losses and/or sweeping up
of jet matter \citep{2007MNRAS.382..526P, 2009ApJ...692.1374B, 2009ApJ...699.1274B}. 

\section{Recollimation Scenario}
Under this framework, the broadband emission is expected to originate from a compact region, compressed by 
the recollimation shock, and this focusing effect may also to be associated with the deceleration
of the jet \citep{2007MNRAS.382..526P, 2009ApJ...699.1274B}. Based on these considerations, we 
assume the emission region to be a spherical 
blob of radius $R^\prime$, moving down the jet with bulk Lorentz factor evolving with time ($t$) 
as
\begin{align}
\label{eq:gammaevol}
{\Gamma}(t) = {\Gamma}_0+{\Gamma}_1 e^{-(t-t_0)/{\tau}}
\end{align}
where ${\Gamma}_{max}={\Gamma}_0+{\Gamma}_1$ is the maximum Lorentz factor corresponding to an initial time
$t_0$, ${\Gamma}_0$ is the terminal Lorentz factor and $1/{\tau}$ is the decay coefficient. A broken power-law electron 
distribution \citep{2008MNRAS.388L..49S} is continuously fed into the emission region and the electrons lose their 
energy through synchrotron, SSC and EC processes. The deceleration of the jet will in turn reduce the Lorentz boosting of 
the external photon energy density, resulting in a time dependent energy loss rate. Besides these radiative losses, the 
electrons also leave the emission region at some characteristic timescale $t_{esc}$. Hence the kinetic equation describing 
the evolution of the electron number density, $n^\prime(\gamma^\prime,t^\prime)$, in the emission region will be
\begin{align}
\label{eq:arkin}
{\partial n^\prime(\gamma^\prime,t^\prime) \over \partial t^\prime} = 
{\partial  \over \partial \gamma^\prime} \left[P(\gamma^\prime, t^\prime) 
	n^\prime(\gamma^\prime,t^\prime)\right] 
	- \frac{n^\prime(\gamma^\prime,t^\prime)}{t_{esc}} + Q (\gamma^\prime,t^\prime) 
 \end{align}
 where $\gamma$ is the electron Lorentz factor, $P(\gamma^\prime, t^\prime)$ is the time dependent total 
 energy loss rate due to synchrotron, SSC and EC processes and $Q(\gamma^\prime,t^\prime)$ is the particle
 injection rate. The radiative energy loss rate, $P(\gamma^\prime, t^\prime)$, due to these emission processes 
can be expressed as \citep{1986rpa..book.....R, 1995ApJ...446L..63D}
\begin{align}
	P(\gamma^\prime, t^\prime)= \frac{4 c \sigma_T}{3 m_e c^2} \gamma^{\prime 2} (u_B^\prime+u_{syn}^\prime+ u_{ir}^\prime)
\end{align}
 Here $u_B^\prime$ is the magnetic field energy density and $u_{syn}^\prime$ is the synchrotron photon energy density.
The particle injection rate, $Q (\gamma^\prime,t^\prime)$, is given by
\begin{align}
\label{eq:broken}
Q(\gamma^\prime,t^\prime) d\gamma^\prime =\left\{
\begin{array}{l l}
	K(t^\prime) \gamma^{\prime -p} d \gamma^\prime; & \quad \gamma_{min}^\prime< \gamma^\prime< \gamma_b^\prime \\
	K(t^\prime) \gamma_b^{\prime q-p} \gamma^{\prime -q} d \gamma^\prime; & \quad \gamma_b^\prime< \gamma^\prime< \gamma_{max}^\prime\\
  \end{array} \right.
\end{align}
where $\gamma_b^\prime$ is the electron Lorentz factor corresponding to the break energy and the spectral indices 
before and after $\gamma_b^\prime$ are given by $p$ and $q$ respectively. The time dependent normalization 
factor, $K(t^\prime)$, is obtained from 
the rate of loss of the bulk energy of the jet, assuming the jet inertia is dominated by cold protons, 
\begin{align}
	K(t^\prime) \approx \left(\frac{m_p\,N_p}{m_e U_e^\star V^\prime}\right) \frac{d\Gamma}{dt}
\end{align}
where $m_p$ is the mass of a proton, $N_p$ is the number of protons occupying the jet volume ($V_p$), and
$V^\prime$ is the proper volume of the emission region and $U_e^\star \approx \gamma_{min}^{\prime 1-p}/(p-1)$.

Equation (\ref{eq:arkin}) cannot be solved analytically due to the form of time dependence in the loss 
term, $P(\gamma^\prime,t^\prime)$, and hence it has been treated numerically following a finite difference scheme prescribed by 
\citet{1970JCoPh...6....1C}. The resultant photon spectrum due to various emission processes, at any instant, can be obtained by convolving the 
particle distribution $n^\prime(\gamma^\prime,t^\prime)$ with the corresponding single particle emissivity.
If the jet makes an angle $\theta$ with respect to the line of sight, then the observer will measure a Doppler boosted 
flux ($F_{obs}$) at time $t_{obs}[\,=\,t(1+z)]$, which after cosmological corrections will be \citep{1984RvMP...56..255B} 
\begin{align} 
\label{eq:trans}
F_{obs}(\nu_{obs}, t_{obs})= \frac{\delta^3(t) (1+z)}{d_L^2} V^\prime \epsilon^\prime\left(\frac{(1+z)}{\delta(t)}\nu_{obs},\frac{\delta(t)}{(1+z)}t_{obs}
\right)
\end{align}
Here $\nu_{obs}$ is the observed photon frequency and $\epsilon^\prime$ is the source emissivity. The deceleration of the 
jet is also associated with the opening of the emission cone with semi vertical angle of $\sim 1/\Gamma(t)$. 
Hence, the time dependence of the observed spectrum is a manifestation of the 
energy loss rate, loss of jet inertia, Doppler boosting and opening angle of the emission cone. 

\section{Results and Discussion}\label{sec:resndis}
The one day binned $\gamma$-ray light-curve of PKS 1222+216 in three different energy bands (Fig. \ref{fig:lc}), during the flare, 
are reproduced by considering EC scattering of IR photons and jet deceleration at the recollimation zone. 
Under this scenario, the observed $\gamma$-ray flare can be interpreted in the following manner: If the jet is misaligned with 
respect to the line of sight of the observer by an angle $\theta > \Gamma_{max}^{-1}$, then this will cause the observer to lie 
outside the emission cone initially, resulting in low observed flux. Now, as the jet decelerates 
and when $\Gamma(t)\lesssim \theta^{-1}$, 
the emission cone widens up causing increase in the observed flux. However, further deceleration of the jet reduces the 
relativistic Doppler boosting and hence the target photon energy density, which eventually lead to the decay of 
the observed flux. We apply this picture to the \emph{Fermi} $\gamma$-ray light-curves and the resultant model light-curves 
along with the observed fluxes are shown in Fig. \ref{fig:lc}. 
The parameters governing the particle injection spectrum and the evolution of bulk Lorentz factor are given in Table 
\ref{tab:parameter}. 

\begin{table*}
\centering
\begin{minipage}{140mm}
\caption{Model parameters} \label{tab:parameter}
\begin{tabular}{ccccccccccccc}
\hline 
p&q& $\gamma_{b}^\prime$ & $B^\prime$ &	$n_p$	& ${\Gamma}_0$ & ${\Gamma}_1$ & $R^{\prime}$ & $\theta$	& $\gamma^{\prime}_{min}$ & $\gamma^{\prime}_{max}$ & ${\tau}$ & $D$\\
\hline \hline
1.15 & 3.0 & $4.2\times 10^3$ & 0.10 & 1.2 & 18 & 39 & $3$ & $2.5$ & 30 & $7\times 10^4$ & $1.3$ & $6$\\
\hline
\end{tabular}
\\ 
Columns: (1) \& (2) Particle spectral indices before and after the break; (3) Lorentz factor of the electron corresponding to break energy; 
(4) Magnetic field (in Gauss); (5) Cold proton number density, $n_p = N_p/V_p$ (in $cm^{-3}$ ); (6) Terminal bulk Lorentz factor; 
(7) Parameter deciding the maximum bulk Lorentz factor, ${\Gamma}_{max} ={\Gamma}_0+{\Gamma}_1$ (8) Emission region size 
($\times10^{14}$ cm); (9) Viewing angle (in degree); 
(10) Minimum Lorentz factor of injected electrons; (11) Maximum Lorentz factor of injected electrons; 
(12) Decay time scale of the bulk Lorentz factor (in days); (13) Jet cross-section at the location of emission region ($\times10^{16}$ cm)\\
 \end{minipage}
 \end{table*}

\begin{figure}
\begin{center}
\includegraphics[scale=1]{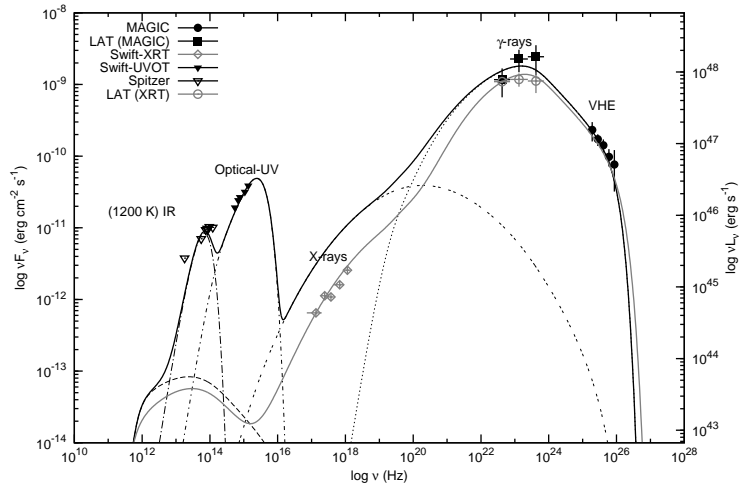}
\end{center}
\caption{ The broadband spectrum of PKS1222+216 at \emph{MAGIC} and XRT observation epoch. The VHE data (black solid circles) 
are obtained from \citet{2011ApJ...730L...8A} and the corresponding LAT fluxes are extracted from two hour (black squares) 
integrated LAT data.  The grey data represents the XRT observation with corresponding LAT fluxes from six hour integrated LAT 
data.The dashed, double dashed  and dotted lines represent the synchrotron, SSC and EC spectra at the time of VHE detection by 
\emph{MAGIC}. The solid black and grey lines represent the total emission from all spectral components at the two epoch 
respectively. The IR-optical-UV data are taken from literature and are reproduced by the torus emission at $1200$ K and multi 
temperature blackbody emission from accretion disk (see \S2 and \S \ref{sec:resndis}).}
 \label{fig:spec}
\end{figure}

The number of parameters governing the spectral and temporal behaviour of the source are large in comparison 
with the information available through the LAT light-curves in three energy 
bands and hence may not be well constrained. However, additional constraint on the parameters can be imposed by 
reproducing the time averaged broadband spectrum of the source, over the \emph{MAGIC} and XRT observation epoch 
(see \S2). In Figure \ref{fig:spec}, we show the resultant spectrum (solid black line) corresponding to the parameters given 
in Table \ref{tab:parameter}, along with the observed fluxes at the time of \emph{MAGIC} detection. The model spectrum 
corresponding to \emph{Swift}--XRT observation period is shown as grey line in Fig. \ref{fig:spec}. The optical/UV and IR fluxes 
are reproduced considering emission from accretion disk and obscuring torus. For disk emission, we adopt a
multi-temperature blackbody with luminosity $L_{UV}\simeq 7\times10^{46}$ erg s$^{-1}$ peaking at 
$3\times10^{15}$ Hz \citep{2002apa..book.....F}. The IR emission from torus is approximated as thermal
radiation, at a temperature of $1200\,K$, originating from a spherical shell of radius $R_{IR}=7\times10^{18}$ cm.
The highest energy of the electron, $\gamma^\prime_{max}$, injected into the emission region (equation (\ref{eq:broken})), 
is assumed to be constant in the present work. However, to interpret the VHE light curve and fit SED properly 
($\sim 10$ min variability) one may need to consider the evolution of $\gamma^\prime_{max}$ also (see Fig. \ref{fig:spec}).

The parameters obtained by reproducing  the $\gamma$-ray light curves of PKS 1222+216 and the broadband 
SED (Table \ref{tab:parameter}), correspond to a jet deceleration with bulk Lorentz factor reducing from 
${\Gamma} \approx 57$ to $18$ over a period of $\approx 7$ days in the AGN frame. We also find that for these 
set of parameters, the flux reaches its maximum $\sim 2.5$ days after the start of particle 
injection, with the corresponding $\Gamma \approx 31$. If we assume that the jet inertia is due to cold 
protons, then the instantaneous jet kinetic power can be expressed as \citep{1997MNRAS.286..415C}
\begin{equation}
	P_{jet}(t) \approx \frac{3}{4} \frac{N_p}{D}(m_pc^2)\Gamma(t)^2 \beta c
\end{equation}
In Figure \ref{fig:lc} (bottom panel), we show the instantaneous jet power during the flaring episode. From this, the average jet power 
during the activity period can be estimated as
\begin{equation}
P_{ave} = \frac{\int P_{jet}(t)dt}{\int dt} \approx  4 \times 10^{44}\; \rm{erg~ s^{-1}} 
\end{equation}
The radiated power during the flare maximum is $P_{rad}\approx 3 \times10^{42}$ erg s$^{-1}$, 
which is much smaller than $P_{ave}$.  Hence, we can conclude that only a small fraction of jet power is utilized as radiation 
in the recollimation zone and most of it is used to launch the jet to larger scales. 

We constrain the size of emission region by considering the VHE variability timescale ($\sim 10$ min, see eq. \ref{eq:VHEvar}) 
and has been kept fixed in the present study. However,
reproduction of spectral and temporal behaviour from such a compact region demands extreme departure from equipartition
with non-thermal particle energy density exceeding the magnetic energy density by $\sim 10^6$. This huge imbalance between
the particle and magnetic pressure may lead to a rapid expansion of the emission region until the pressure balance is achieved.
For an expansion velocity of $v_{exp}^\prime\approx c/3$, near equipartition is achieved towards the end of the LAT flare when 
the emission region would have expanded to a size of $\sim 10^{16} cm$, comparable to the jet cross-section. The adiabatic 
cooling timescale ($t_{ad}^\prime$) associated with the expansion of the emission region of size $R^\prime(t_{obs}^\prime)$ at any 
instant $t_{obs}^\prime[=\left(\frac{\delta}{1+z} \right) t_{obs}]$ is
\begin{equation}
 t_{ad}^\prime \approx \frac{R^\prime(t_{obs}^\prime)}{v_{exp}^\prime}
\end{equation}
where $v_{exp}^\prime$ is the expansion speed. If $R_i^\prime$ is the initial size of the emission region then from 

\begin{equation}
R^\prime(t_{obs}^\prime)=R_i^\prime+v_{exp}^\prime  t_{obs}^\prime
\end{equation}
 we obtain $t_{ad}^\prime =\frac{R_i^\prime}{v_{exp}^\prime}+t_{obs}^\prime > t_{obs}^\prime$.
However, the radiative cooling timescale is much shorter
compared to the flare duration (\S2) and hence the effect of adiabatic cooling on particle
evolution will be negligible. 

A decelerating jet model was earlier proposed by \cite{2003ApJ...594L..27G}, in order to explain the TeV emission from the BL Lac 
objects Mrk 421 and Mrk 501 with reasonable jet Lorentz factors. They considered the inverse Compton scattering of synchrotron 
photons from a slower moving part of the jet by the relativistic electrons present in the faster moving part (upstream). 
In such a situation, the target photons will be beamed in the frame of the faster moving part of the jet and get scattered to VHE by the 
relativistic electrons. \cite{2009ApJ...692.1374B}  also proposed a decelerating jet model to explain the decay of the optical 
light-curve of 3C 279 during a flare observed in 2006. Their model is analogous to the relativistic blast-wave model for 
$\gamma$-ray bursts where they considered radiative energy losses and radiative drag being the main reasons for jet deceleration. 
However in the present work, we perform a detailed study of the evolution of particle distribution under a decelerating jet 
scenario, involving different loss processes. Temporal behaviour of the spectrum is a reflection of time dependent particle injection rate,
resulting from the loss of jet power, and relativistic effects related to jet deceleration.
Recently, \cite{2013MNRAS.429.1189P} developed a model where they considered jet acceleration from a Poynting flux
dominated region followed by deceleration after attaining the terminal bulk Lorentz factor. Using this model they explained the
broadband SED of PKS\,0227-369. In the present work, we have considered emission from a location where jet is already
kinetic energy dominated.

Interpretation of blazar flare as a result of difference in bulk Lorentz factor was earlier used by \cite{2001ApJ...553..683H} to 
explain the multi-wavelength spectra of blazar 3C 279 during $\gamma$-ray high state in 1999 and 2000. They fitted 
the broadband spectra at different epochs by varying the bulk Lorentz factor of the jet accompanied by a change in 
the spectral shape of the electron distribution. The emission region was within the BLR clouds and hence significant 
contribution of the external Compton radiation was obtained by scattering the BLR photons. 
Here, we model a single $\gamma$-ray flare by evolving the jet Lorentz factor gradually. Simultaneous detection of the source at VHE also 
suggests that the emission region lies outside the BLR clouds and hence the external Compton radiation will be 
dominated by the scattering of IR photons from the torus. Our model
is similar to the blob-in-jet model used by \cite{2011A&A...534A..86T} to explain the broadband SED
of PKS 1222+216. However, in addition to reproducing the broadband SED, we explain the LAT light curves by considering the 
jet dynamics at the recollimation zone. Blob-in-jet model, incorporating only SSC process, has also
been used to model the VHE $\gamma$-ray light-curve and SED of TeV BL Lac blazar PKS 2155-304 during a flare in 
2006 \citep{2012A&A...539A.149H}.

\section{Conclusions}\label{sec:conclude}
Detection of VHE $\gamma$-rays with a variability timescale of $\sim 10$ min suggest a very compact emission region
located beyond BLR. Existence of such compact emission region, on parsec scale, requires strong convergence of the jet flow, 
suggesting recollimation as one of the possible mechanism. In this paper, the variation in the $\gamma$-ray spectrum of 
PKS 1222+216, observed by {\emph Fermi}-LAT, is explained considering the jet dynamics at recollimation zone. 
Besides providing a compact emission region due to compression of jet matter by recollimation shock, study
of jet dynamics at recollimation zone suggests deceleration of jet flow. We adapt this scenario
to reproduce the daily binned $\gamma$-ray light-curves observed in three different energy bands. The parameters
governing the model are further constrained by reproducing the simultaneous/contemporaneous broadband SED of the source
during the flare. The inferred values of the bulk Lorentz factor ($\Gamma$) and Doppler factor ($\delta$)
are consistent with the ones estimated through radio and $\gamma$-ray  studies of LAT bright blazars \citep{2012IJMPS...8..163H}. 

Due to the lack of simultaneous observations at radio/optical (synchrotron) and X-ray (SSC) 
energies during the flare, the light-curves at these energies cannot be compared with the present 
model. Future simultaneous multi-wavelength observations  of the source at energies 
covering from radio-to-$\gamma$-ray can verify the present model and can be used to impose more stringent constraints 
on the parameters involved. The spectral evolution of the source at these energies 
will also help us in studying the effect of magnetic field and the underlying particle distribution during a flare. For instance, 
variation of magnetic field in the emission region along with the bulk Lorentz factor during a flare will lead to additional 
increase/decrease in the synchrotron and SSC fluxes along with a shift in their peak frequencies. Similarly, a change in the
spectral index can throw light on the basic particle acceleration mechanism. These in turn will help us in understanding 
the energetics and dynamics of the AGN jets.

Authors thank the anonymous referees for their useful comments and suggestions.  PK thanks
K Nalewajko for clarification on some of the calculations in his paper \citep{2012MNRAS.425.2519N}. 
SS acknowledges Ranjeev Misra for useful discussions. 
This research has made use of data obtained from High Energy Astrophysics Science Archive Research Center 
(HEASARC), maintained by NASA's Goddard Space Flight Center and NASA/IPAC Extra-galactic Database (NED) which is operated by the Jet 
Propulsion Laboratory, California Institute of Technology, under contract with the NASA. 

\begin{thebibliography}{}

\bibitem[\protect\citeauthoryear{Albert et al.}{2007}]{2007ApJ...669..862A} 
Albert J., et al., 2007, ApJ, 669, 862 

\bibitem[\protect\citeauthoryear{Aleksi{\'c} et 
al.}{2011}]{2011ApJ...730L...8A} Aleksi{\'c} J., et al., 2011, ApJ, 730, L8 

\bibitem[\protect\citeauthoryear{Atwood et al.}{2009}]{2009ApJ...697.1071A} 
Atwood W.~B., et al., 2009, ApJ, 697, 1071

\bibitem[\protect\citeauthoryear{Begelman, Blandford, 
\& Rees}{1984}]{1984RvMP...56..255B} Begelman M.~C., Blandford R.~D., Rees M.~J., 1984, RvMP, 56, 255

\bibitem[\protect\citeauthoryear{Begelman 
\& Sikora}{1987}]{1987ApJ...322..650B} Begelman M.~C., Sikora M., 1987, ApJ, 322, 650

\bibitem[\protect\citeauthoryear{B{\l}a{\.z}ejowski et 
al.}{2000}]{2000ApJ...545..107B} B{\l}a{\.z}ejowski M., Sikora M., Moderski 
R., Madejski G.~M., 2000, ApJ, 545, 107

\bibitem[\protect\citeauthoryear{B{\"o}ttcher 
\& Principe}{2009}]{2009ApJ...692.1374B} B{\"o}ttcher M., Principe D., 2009, ApJ, 692, 1374 

\bibitem[\protect\citeauthoryear{B{\"o}ttcher, Reimer, 
\& Marscher}{2009}]{2009ApJ...703.1168B} B{\"o}ttcher M., Reimer A., Marscher A.~P., 2009, ApJ, 703, 1168

\bibitem[\protect\citeauthoryear{Boettcher, Reuter, 
\& Lesch}{1997}]{1997A&A...326L..33B} Boettcher M., Reuter H.-P., Lesch H., 1997, A\&A, 326, L33

\bibitem[\protect\citeauthoryear{Bromberg 
\& Levinson}{2009}]{2009ApJ...699.1274B} Bromberg O., Levinson A., 2009, ApJ, 699, 1274

\bibitem[\protect\citeauthoryear{Bulgarelli et 
al.}{2010}]{2010ATel.2641....1B} Bulgarelli A., et al., 2010, ATel, 2641, 1

\bibitem[\protect\citeauthoryear{Carrasco et 
al.}{2010}]{2010ATel.2626....1C} Carrasco L., Carrami{\~n}ana A., Recillas 
E., Porras A., Mayya D.~Y., 2010, ATel, 2626, 1

\bibitem[\protect\citeauthoryear{Celotti, Padovani, 
\& Ghisellini}{1997}]{1997MNRAS.286..415C} Celotti A., Padovani P., Ghisellini G., 1997, MNRAS, 286, 415

\bibitem[Chiang 
\& Dermer(1999)]{1999ApJ...512..699C} Chiang, J., \& Dermer, C.~D.\ 1999, \apj, 512, 699 

\bibitem[\protect\citeauthoryear{Chang 
\& Cooper}{1970}]{1970JCoPh...6....1C} Chang J.~S., Cooper G., 1970, JCoPh, 6, 1 

\bibitem[\protect\citeauthoryear{Dermer}{1995}]{1995ApJ...446L..63D} Dermer 
C.~D., 1995, ApJ, 446, L63

\bibitem[\protect\citeauthoryear{Dermer 
\& Menon}{2009}]{2009herb.book.....D} Dermer C.~D., Menon G., 2009, High Energy Radiation from Black Holes, Princeton
University Press, NJ

\bibitem[\protect\citeauthoryear{Dermer, Schlickeiser, 
\& Mastichiadis}{1992}]{1992A&A...256L..27D} Dermer C.~D., Schlickeiser R., Mastichiadis A., 1992, A\&A, 256, L27

\bibitem[\protect\citeauthoryear{Dermer 
\& Schlickeiser}{1993}]{1993ApJ...416..458D} Dermer C.~D., Schlickeiser R., 1993, ApJ, 416, 458

\bibitem[\protect\citeauthoryear{Dondi 
\& Ghisellini}{1995}]{1995MNRAS.273..583D} Dondi L., Ghisellini G., 1995, MNRAS, 273, 583 

\bibitem[\protect\citeauthoryear{Fan et al.}{2008}]{2008PASJ...60..707F} 
Fan J.-H., et al., 2008, PASJ, 60, 707

\bibitem[\protect\citeauthoryear{Frank, King, 
\& Raine}{2002}]{2002apa..book.....F} Frank J., King A., Raine D.~J., 2002, Accretion Power in Astrophysics, Cambridge 
University Press, UK

\bibitem[\protect\citeauthoryear{Georganopoulos 
\& Kazanas}{2003}]{2003ApJ...594L..27G} Georganopoulos M., Kazanas D., 2003, ApJ, 594, L27

\bibitem[\protect\citeauthoryear{Ghisellini 
\& Maraschi}{1989}]{1989ApJ...340..181G} Ghisellini G., Maraschi L., 1989, ApJ, 340, 181

\bibitem[\protect\citeauthoryear{Ghisellini 
\& Madau}{1996}]{1996MNRAS.280...67G} Ghisellini G., Madau P., 1996, MNRAS, 280, 67

\bibitem[\protect\citeauthoryear{Ghisellini 
\& Tavecchio}{2008}]{2008MNRAS.387.1669G} Ghisellini G., Tavecchio F., 2008, MNRAS, 387, 1669

\bibitem[\protect\citeauthoryear{Ghisellini 
\& Tavecchio}{2009}]{2009MNRAS.397..985G} Ghisellini G., Tavecchio F., 2009, MNRAS, 397, 985 

\bibitem[\protect\citeauthoryear{Hartman et 
al.}{2001}]{2001ApJ...553..683H} Hartman R.~C., et al., 2001, ApJ, 553, 683

\bibitem[\protect\citeauthoryear{H.E.S.S.~Collaboration et 
al.}{2012}]{2012A&A...539A.149H} H.E.S.S.~Collaboration, et al., 2012, A\&A, 539, A149

\bibitem[\protect\citeauthoryear{Homan}{2012}]{2012IJMPS...8..163H} Homan 
D.~C., 2012, IJMPS, 8, 163

\bibitem[\protect\citeauthoryear{Kalberla et 
al.}{2005}]{2005A&A...440..775K} Kalberla P.~M.~W., Burton W.~B., Hartmann D., Arnal E.~M., Bajaja E., Morras R., P{\"o}ppel W.~G.~L., 2005, A\&A, 440, 775

\bibitem[\protect\citeauthoryear{Kataoka et 
al.}{1999}]{1999ApJ...514..138K} Kataoka J., et al., 1999, ApJ, 514, 138

\bibitem[\protect\citeauthoryear{Konigl}{1981}]{1981ApJ...243..700K} Konigl 
A., 1981, ApJ, 243, 700

\bibitem[\protect\citeauthoryear{Krawczynski}{2004}]{2004NewAR..48..367K} 
Krawczynski H., 2004, NewAR, 48, 367

\bibitem[\protect\citeauthoryear{Kushwaha, Sahayanathan, 
\& Singh}{2013}]{2013MNRAS.433.2380K} Kushwaha P., Sahayanathan S., Singh K.~P., 2013, MNRAS, 433, 2380 

\bibitem[\protect\citeauthoryear{MAGIC Collaboration et 
al.}{2008}]{2008Sci...320.1752M} MAGIC Collaboration, et al., 2008, Sci, 
320, 1752

\bibitem[\protect\citeauthoryear{Malmrose et 
al.}{2011}]{2011ApJ...732..116M} Malmrose M.~P., Marscher A.~P., Jorstad 
S.~G., Nikutta R., Elitzur M., 2011, ApJ, 732, 116 

\bibitem[Marscher(1999)]{1999APh....11...19M} Marscher, A.~P.\ 1999, 
Astroparticle Physics, 11, 19

\bibitem[\protect\citeauthoryear{Marscher 
\& Gear}{1985}]{1985ApJ...298..114M} Marscher A.~P., Gear W.~K., 1985, ApJ, 298, 114

\bibitem[\protect\citeauthoryear{Mattox et al.}{1996}]{1996ApJ...461..396M} 
Mattox J.~R., et al., 1996, ApJ, 461, 396

\bibitem[\protect\citeauthoryear{Melia 
\& Konigl}{1989}]{1989ApJ...340..162M} Melia F., Konigl A., 1989, ApJ, 340, 162

\bibitem[\protect\citeauthoryear{Moretti et 
al.}{2005}]{2005SPIE.5898..360M} Moretti A., et al., 2005, SPIE, 5898, 360

\bibitem[\protect\citeauthoryear{Mose Mariotti}{2010}]{2010ATel.2684....1M} 
Mose Mariotti M., 2010, ATel, 2684, 1

\bibitem[\protect\citeauthoryear{Nalewajko et 
al.}{2012}]{2012MNRAS.425.2519N} Nalewajko K., Begelman M.~C., Cerutti B., 
Uzdensky D.~A., Sikora M., 2012, MNRAS, 425, 2519

\bibitem[\protect\citeauthoryear{Nolan et al.}{2012}]{2012ApJS..199...31N} 
Nolan P.~L., et al., 2012, ApJS, 199, 31

\bibitem[\protect\citeauthoryear{Perucho 
\& Mart{\'{\i}}}{2007}]{2007MNRAS.382..526P} Perucho M., Mart{\'{\i}} J.~M., 2007, MNRAS, 382, 526 

\bibitem[\protect\citeauthoryear{Potter 
\& Cotter}{2013}]{2013MNRAS.429.1189P} Potter W.~J., Cotter G., 2013, MNRAS, 429, 1189 

\bibitem[\protect\citeauthoryear{Pushkarev et 
al.}{2009}]{2009A&A...507L..33P} Pushkarev A.~B., Kovalev Y.~Y., Lister M.~L., Savolainen T., 2009, A\&A, 507, L33

\bibitem[\protect\citeauthoryear{Rybicki 
\& Lightman}{1986}]{1986rpa..book.....R} Rybicki G.~B., Lightman A.~P., 1986, Radiative Processes in Astrophysics, Wiley, New York

\bibitem[\protect\citeauthoryear{Sahayanathan}{2008}]{2008MNRAS.388L..49S} 
Sahayanathan S., 2008, MNRAS, 388, L49 

\bibitem[\protect\citeauthoryear{Sahayanathan 
\& Godambe}{2012}]{2012MNRAS.419.1660S} Sahayanathan S., Godambe S., 2012, MNRAS, 419, 1660

\bibitem[\protect\citeauthoryear{Sambruna}{2000}]{2000AIPC..515...19S} 
Sambruna R.~M., 2000, AIPC, 515, 19

\bibitem[\protect\citeauthoryear{Sikora, Begelman, 
\& Rees}{1994}]{1994ApJ...421..153S} Sikora M., Begelman M.~C., Rees M.~J., 1994, ApJ, 421, 153

\bibitem[\protect\citeauthoryear{Tanaka et al.}{2011}]{2011ApJ...733...19T} 
Tanaka Y.~T., et al., 2011, ApJ, 733, 19 

\bibitem[\protect\citeauthoryear{Tavecchio et 
al.}{2011}]{2011A&A...534A..86T} Tavecchio F., Becerra-Gonzalez J., Ghisellini G., Stamerra A., Bonnoli G., Foschini L., Maraschi L., 2011, A\&A, 534, A86 

\bibitem[\protect\citeauthoryear{Urry 
\& Padovani}{1995}]{1995PASP..107..803U} Urry C.~M., Padovani P., 1995, PASP, 107, 803

\bibitem[\protect\citeauthoryear{Wagner 
\& H.E.S.S.~Collaboration}{2010}]{2010HEAD...11.2706W} Wagner S.~J., H.E.S.S.~Collaboration, 2010, HEAD, 11, \#27.06 

\bibitem[\protect\citeauthoryear{Wehrle et al.}{1998}]{1998ApJ...497..178W} 
Wehrle A.~E., et al., 1998, ApJ, 497, 178

\end{thebibliography}
\end{document}